\documentclass[12pt]{article}
\usepackage[dvips]{graphicx}
\usepackage{latexsym}
\usepackage{hyperref}
\parskip=\medskipamount

\newcommand {\cA}{{\cal A}}

\newcommand {\cD}{{\cal D}}

\newcommand {\cF}{{\cal F}}

\newcommand {\cN}{{\cal N}}

\newcommand {\cY}{{\cal Y}}

\def\a{\alpha}

\def\b{\beta}

\def\d{\delta}
\def\e{\epsilon}

\def\g{\gamma}
\def\G{\Gamma}

\def\m{\mu}
\def\n{\nu}

\def\p{\pi}

\def\r{\rho}
\def\s{\sigma}
\def\t{\tau}

\def\z{\zeta}
\def\D{\Delta}

\def\tr{{\rm tr}}
\def\rd{{\rm d}}
\def\ri{i}
\def\re{{\rm e}}

\newcommand{\sect}[1]{\setcounter{equation}{0}\section{#1}}

\newcommand{\be}{\begin{equation}}
\newcommand{\ee}{\end{equation}}
\newcommand{\bea}{\begin{eqnarray}}
\newcommand{\eea}{\end{eqnarray}}

\newcommand{\iu}{\underline{i}}
\newcommand{\ju}{\underline{j}}

\newcommand{\Iu}{\underline{I}}

\begin{document}

\begin{titlepage}
\thispagestyle{empty}


\begin{center}
{\Large \bf
The SU(N) Matrix Model at Two Loops}
\end{center}

\begin{center}
{\large D. Grasso and   I.N. McArthur}\\
\vspace{2mm}

\footnotesize{
{\it School of Physics, The University of Western Australia\\
Crawley, W.A. 6009, Australia}
} \\
{\tt  grasso@physics.uwa.edu.au},~
{\tt mcarthur@physics.uwa.edu.au} \\
\end{center}
\vspace{5mm}

\begin{abstract}
\baselineskip=14pt

Multi-loop calculations of the effective action for the matrix model are important for carrying out tests of the conjectured relationship of the matrix model to the low energy description of M-theory. In particular, comparison with $N$-graviton scattering amplitudes in eleven-dimensional supergravity requires the calculation of the effective action for the matrix model with gauge group $SU(N).$ A framework for carrying out such calculations at two loops is established in this paper. The two-loop effective action is explicitly computed for a background corresponding to the scattering of a single D0-brane from a stack of $N-1$ D0-branes, and the results are shown to agree with known results in the case $N=2.$

\end{abstract}

\vfill
\end{titlepage}

\newpage
\setcounter{page}{1}

\renewcommand{\thefootnote}{\arabic{footnote}}
\setcounter{footnote}{0}
\sect{Introduction}
The matrix model  is a quantum mechanical theory resulting from the dimensional reduction of ten-dimensional supersymmetric Yang-Mills theory to 1+0 dimensions \cite{CH}.  In 1996,  Banks, Fischler, Shenker and Susskind (BFSS) conjectured that the dynamical degrees of freedom of M-theory in the infinite momentum frame correspond to the large $N$ limit of a system of $N$ D0-branes \cite{Banks:1996vh}. The $SU(N)$ matrix model describes the low energy dynamics of a system of $N$ D0-branes \cite{Witten:1995im}, with interactions between D0-branes arising from quantum loop corrections \cite{Douglas:1996yp}, so this conjecture resulted in an explosion of activity related to the matrix model (for a review, see \cite{Taylor:2001vb}). The matrix model for finite $N$ is conjectured to be related to the discrete light-front quantized sector of M-theory \cite{Susskind:1997cw,Seiberg:1997ad,Sen:1998we}.

Tests of the BFSS conjecture require comparison of graviton scattering amplitudes in  eleven-dimensional supergravity with loop calculations in the matrix model. The bulk of the tests so far carried out pertain to two graviton scattering, for which the gauge group for the matrix model can be chosen to be $SU(2).$ The exception is the work of Okawa and Yoneya \cite{Okawa:1998pz,Okawa:1998qk}, in which three graviton scattering is considered.  General $N$-graviton scattering amplitudes require loop calculations in the matrix model with a gauge group $SU(N),$ and this paper provides a framework for such calculations.

Agreement between classical graviton scattering amplitudes and matrix model predictions has been impressive. The one-loop effective action for the SU(2) matrix model successfully reproduces a $\frac{v^4}{r^7}$ term in the scattering potential \cite{Douglas:1996yp,Banks:1996vh}.  Becker and Becker computed the leading two-loop contribution to the effective action \cite{Becker:1997wh}, and showed that the coefficient of the $\frac{v^4}{r^{10}}$ term is zero, in agreement with the BFSS conjecture  that the $O(v^4)$ terms in the potential should not be renormalized beyond one loop. This calculation was extended in \cite{Becker:1997xw} and yielded agreement with supergravity predictions for the coefficient of the $\frac{v^6}{r^{14}}$ term in the scattering potential.\footnote{The  full two-loop effective action for the $SU(2)$ matrix model was later presented in \cite{Becker:1998gp}.}

The equivalence between  linearized gravitational interactions and the one-loop matrix model has been established in general for a pair of bosonic sources \cite{Taylor:1997jb}. Spin-dependent effects in the two graviton interaction have been shown to be reproduced by   matrix model in the presence  of a fermionic background at one-loop \cite{Harvey:1998ic,Morales:1998hk,Morales:1998dz,Kraus:1998st,Barrio:1998hr,Plefka:1998ed,McArthur:1998gs,Plefka:1998xu,Taylor:1999tv,Hyun:1999hm,Hyun:1999qd,Nicolai:2000ht}. The extent to which this agreement is a consequence of  supersymmetric non-renormalization theorems has been investigated in \cite{Paban:1998ea,Paban:1998qy,Hyun:1999hf,Okawa:1999hi,Kazama:2000ar}. Supersymmetric nonrenormalization theorems are expected to be much less restrictive at two-loops for gauge group $SU(N)$ with $N>3$ \cite{Sethi:1999qv}.

The situation for {\it quantum} corrections to eleven-dimensional graviton scattering amplitudes is less clear. There are indications that  the $\frac{v^8}{r^{18}}$ term in the scattering potential, arising from quantum corrections on the supergravity side \cite{Susskind,Berglund:1997hj}, is not correctly reproduced in the two-loop effective action for the matrix model \cite{Keski-Vakkuri:1997wr,Becker:1997cp,Helling:1999js}. It is possible that quantum effects in M-theory are only reproduced in the large $N$ limit of the matrix model  \cite{Taylor:2001vb}.

In this paper, we conduct a two-loop calculation of the effective action for the matrix model in the case where the gauge group is $SU(N).$ The background chosen is appropriate for consideration of the scattering of a single D0-brane from a stack on $N$-1 coincident D0-branes. However, the results have wider applicability. The two-loop Feynman diagrams computed  form a subclass of those required for the  study of two-loop contributions to the scattering potential for $N$ D0-branes in  more general configurations.  

In the case $N=2,$ the results are shown to be in agreement with the results of the Beckers \cite{Becker:1997wh,Becker:1998gp}, although this requires the use of some nontrivial identities for hypergeometric functions. This agreement provides an important check of the validity of general results on the group theoretic structure of two-loop graphs which were developed in \cite{Kuzenko:2003wu}. These group theoretic results have already been used in two-loop calculations of the effective action  for ${\cal N} = 4$ supersymmetric Yang-Mills theory with gauge group $SU(N)$ in four dimensions \cite{Kuzenko:2003wu,Kuzenko:2004ma,Kuzenko:2004sv}.  An independent check of the group theoretic results in \cite{Kuzenko:2003wu} is useful in light of the fact that some of the outcomes in \cite{Kuzenko:2003wu,Kuzenko:2004ma,Kuzenko:2004sv} are somewhat unexpected.

The original two-loop calculations for gauge group $SU(2)$ \cite{Becker:1997wh,Becker:1998gp} made use of dimensional regularization to deal with an apparent divergence. The presence of divergences would be somewhat surprising given that the matrix model is a quantum mechanical theory possessing maximal supersymmetry. In this paper, no divergences are encountered in the two-loop calculations. 

The outline of the paper is as follows. In section 2, the matrix model with gauge group $SU(N)$ is reviewed in the context of background field quantization. A specific background relevant to the scattering of a single D0-brane from a stack of $N$-1 D0-branes is chosen in section 3. The general structure of the Feynman diagrams contributing to the two-loop effective action is derived in section 4. In particular, we take advantage of the ten-dimensional origin of the matrix model in the organization of the calculations.  These results lay the basis for  two-loop calculations in the $SU(N)$ matrix model for arbitrary backgrounds. Sections 5 to 9 provide details of the evaluation of these Feynman diagrams in the specific background of interest in this paper. Comparison with the results of  \cite{Becker:1998gp} for $SU(2)$ is carried out in section 10, followed by the conclusion. Details of the evaluation of the heat kernels used to obtain exact Green's functions in an $SU(N)$ background are contained in an appendix, which also contains a discussion of the difference between calculating quantum corrections in ten dimensions and dimesionally reducing as opposed to dimensionally reducing first and then performing quantum calculations. A second appendix establishes the equivalence of two Feynman diagrams.

\sect{Structure of the $SU(N)$ matrix model}
The action for ten-dimensional\footnote{We use the metric $(-, +, +, \cdots , +)$ in this paper. The ten-dimensional gamma matrices satisfy $\{ \Gamma^{\m}, \Gamma^{\n} \} = -2 \, \eta^{\m \n}\, {\bf 1}. $ The trace of $SU(N)$ generators in the fundamental representation is normalized so that ${\rm tr}_{\rm F} (T_a T_b)= \delta_{ab}.$} $\cN = 1$ supersymmetric Yang-Mills theory is
\be
 S = \int \rd^{10}x \, {\rm  tr}_{\rm F}\left(-\, \frac14 \,  \cF^{\mu \nu} \cF_{\mu \nu} + \frac{\ri}{2} \, \bar{\Psi} \Gamma^{\m} \cD_{\m} \Psi \right),
\label{S10}
\ee
where all fields are Lie algebra-valued in the fundamental representation of $SU(N),$ $\Psi$ is a sixteen-component Majorana-Weyl spinor ($\bar{\Psi} = \Psi^T C$), 
$\cF_{\m \n} = \partial_{\m} \cA_{\n} - \partial_{\n} \cA_{\m} +\ri \, g \, [\cA_{\m}, \cA_{\n}],$  $ \cD_{\m} \Psi = \partial_{\m} \Psi + \ri \, g \, [\cA_{\mu}, \Psi],$ and $g$ is the coupling constant. The action for the matrix model is obtained by dimensional reduction of this action to $1+0$ dimensions. In particular, upon dimensional reduction,
$$\cA_{\m} \rightarrow ( \cA_0, \cY_i ), \quad i= 1, \cdots 9, $$ where the $\cY_i$ are scalars in $1+0$ dimensions. However, in deriving the Feynman rules for the matrix model, it is convenient to initially maintain the ten-dimensional Lorentz notation, except that  $\int {\rm d}^{10}x \rightarrow \int {\rm d} \t,$ and with the understanding that there are no spatial derivatives in the covariant derivatives. In the background field quantization scheme, the fields are split into background and quantum pieces. For the purposes of this paper, it suffices to set the fermionic component of the background fields to zero, so
$$ \cA_{\m} \rightarrow A_{\m} + a_{\m}, \quad \Psi \rightarrow \psi,$$
where $A_{\m}$ are background fields and $a_{\m}, \psi$ are quantum fields. After inclusion of the gauge-fixing term
$$S_{g.f.} = - \, \frac12 \, \int {\rm d} \t \, {\rm tr}_{\rm F} \left( (D^{\m} a_{\m})^2 \right)$$ 
with $D_{\m} $ the background covariant derivative,
the gauge-fixed matrix model action is 
\bea
S + S_{g.f.} & = & \int {\rm d} \t \, {\rm tr}_{\rm F} \left(  \frac12 a^{\m} D^{\n}D_{\n} a_{\m} + \ri   g  a^{\m} \, [ F_{\m}{}^{\n}, a_{\n} ] - \ri  g  (D_{\m} a_{\n}) \,[a^{\m}, a^{\n} ] \right. \nonumber \\
&+& \left. \frac{g^2}{4}\, [a_{\m}, a_{\n}] \,  [a^{\m}, a^{\n}] + \frac{\ri}{2} \psi \,(C \G)^{\m} \, (D_{\m} \psi + \ri \, g \, [a_{\m}, \psi ]) \right).
\label{Sgf}
\eea
The corresponding ghost action is
\be
S_{ghost} = \int {\rm d} \t \, {\rm tr}_{\rm F} \left( \bar{c} \, D^{\m} (D_{\m} c +\ri g [a_{\m}, c]) \right).
\label{Sg}
\ee

The propagators  are determined by the piece of the action quadratic in the quantum fields. Decomposing a generic Lie algebra-valued field $\phi$ in the form $\phi = \phi^a \, T_a,$ the propagators take the form\footnote{Note that Roman letters from the start of the alphabet are used for gauge indices, Greek letters from the start of the alphabet are used for spinor indices, and Greek letters from the middle of the alphabet are used for vector indices.}
\bea
\langle a_{\m}^a(\t) \, a_{\n}^b(\t') \rangle &=&    - \, \ri \, G_{\m \n}^{ab} (\t,\t')  \label{aprop}\\
\langle \psi^a_{\a}(\t) \,\psi^{b}_{ \b}(\t') \rangle &=& - \, \frac{\ri }{2}\, G^a_{\a}{}^{b}_{ \b}(\t, \t')  \label{psiprop} \\
\langle c^a(\t) \, \bar{c}^{\, b}(\t') \rangle &=&   - \, \ri \,G^{ab}(\t,\t'), \label{cprop}
\eea
where the Green's functions (with gauge indices suppressed) are defined by 
\bea
\left( D^{\s}D_{\s} \, \d^{\m}{}_{\r}  + 2 \ri  g  F^{\m}{}_{\r}(\t) \right) \, G^{\r}{}_{ \n} (\t,\t') &=& - \, \d(\t,\t') \, \d^{\m}{}_{\n} \label{aGreen} \\
\frac{\ri}{2} \, ( \Gamma^{\mu} )_{\a}{}^{\g} \, D_{\mu} \, G_{\g \b}(\t, \t') &=&  - \, \d(\t,\t') \, (C^{-1})_{\a \b} \label{psiGreen} \\
D^{\m}D_{\m} \, G(\t,\t') &=& - \, \d(\t,\t'). \label{cGreen}
\eea

It is convenient to express the Green's functions in terms of heat kernels. For the Green's function (\ref{cGreen}), 
\be
 G(\t,\t') = \ri \, \int_0^{\infty}  {\rm d}s \, K(\t,\t';s)
\label{kernel}
\ee
where the heat kernel $K(\t,\t';s)$ satisfies the differential equation 
\be
- \, \ri \frac{\rd }{\rd s} \,  K(\t,\t';s) = D^{\m}D_{\m} \, K(\t,\t';s)
\ee
with the boundary condition $\lim_{s \rightarrow 0} K(\t,\t';s) = \d(\t, \t'). $
We compute the Green's functions in the case where the background fields $F_{\m \n}$ (with components $F_{0 i} = D_{\t }Y_i, $ $F_{ij} = \ri \, g \,  [Y_i, Y_j]$) are constant. In this case, the Green's function (\ref{aGreen}) takes the form
\be
G^{\m}{}_{ \n} (\t,\t') =  \ri \, \int_0^{\infty} {\rm d}s \, \left( \re^{-2  s g F} \right)^{\m}{}_{ \n} \,K(\t,\t';s),
\label{factoredvector}
\ee
while the Green's function (\ref{psiGreen}) takes the form
\be
G_{\a \b} (\t,\t') =  -\, 2 \,\int_0^{\infty} {\rm d}s \,  (\Gamma^{\m})_{\a}{}^{ \g}  D_{\m} \, \left( \re^{\frac{s}{4} [\Gamma^{\n}, \Gamma^{\r}] g F_{\n \r} } C^{-1} \right)_{\g \b}  \,K(\t,\t';s).
\ee

As detailed in Appendix A, for constant background field strength, the kernel $K(\t, \t'; s)$ is (with the notation $\vec{F} = D_{\t} \vec{Y}, F^2 = \vec{F}.\vec{F}$ and $Y^2 = \vec{Y}.\vec{Y}$)
\bea
 K(\t, \t'; s) &=& \left( \frac{\ri }{4 \pi s} \right)^{\frac12}  \re^{- \ri  s g^2 (Y^2 - F^2 (\frac{\vec{Y}.  \vec{F}}{F^2})^2)} \, \left( \frac{2 g s F}{\sinh 2gsF} \right)^{\frac12}  \, \exp \left\{ - \, \ri g F \frac{ (\t-\t')^2}{2} \coth 2gsF \quad \quad \right.  \nonumber \\
&+& \left.   \ri gF \,  \left( \frac{\vec{Y}. \vec{F}}{F^2} \right) \, \left( \t - \t' -  \frac{\vec{Y}. \vec{F}}{F^2} \right) \, \tanh gsF  \right\} \, I(\t,\t').
\label{SUNkernel}
\eea
Here,  $I(\t,\t')$ is the parallel displacement propagator which ensures that the Green's function has the correct gauge transformation properties at $\t$ and $\t'.$ For further details on the parallel displacement propagator and its properties, see   \cite{Kuzenko:2003eb}. With an appropriate choice of background, the kernel (\ref{SUNkernel}) reduces to that presented in \cite{Okawa:1998pz}.  We also require the covariant derivative of the kernel, and again as detailed in Appendix A, the result is 
\be
D_{\t}K(\t,\t';s) = \frac{- \, \ri gF}{\sinh 2 gs F}\,  \left\{  (\t - \t') + (\cosh 2g s F - 1) \, \left( \frac{\vec{Y}. \vec{F}}{F^2} \right) 
\right\} \, K(\t,\t';s).
\label{SUNkernelderiv}
\ee

\sect{Specifying the background}
We wish to consider the matrix model with gauge group $SU(N),$ and make use of the $SU(N)$ conventions in \cite{Kuzenko:2003wu}. Matrices in the fundamental representation carry lower-case Latin indices $i, j, \cdots$ from the middle of the alphabet, with the range $i = 0, 1, \cdots, N-1 \equiv 
0, \iu.$ The Cartan generators are labelled by the index $I = 0, 1, \cdots, N-2 \equiv 0, \Iu.$ In the fundamental representation, the generators of the Cartan-Weyl basis $\{H_I, E_{ij}, i\neq j \} $ are represented by the matrices
\bea
(E_{ij})_{kl} &=& \delta_{ik} \, \delta_{kl}, \nonumber \\
(H_I)_{kl} &=& (N-I)^{-\frac12} \, (N-I-1)^{-\frac12} \{(N-I) \, \delta_{kI} \, \delta_{lI} - \sum_{i=I}^{N-1} \delta_{ki} \, \delta_{li}\},
\eea
and are normalized so that ${\rm tr}_{\rm F} (H_I \, H_J) = \delta_{IJ},$ ${\rm tr}_{\rm F} (E_{ij}\, E_{kl}) = \delta_{il}\, \delta_{jk},$  ${\rm tr}_{\rm F} (H_I \, E_{kl}) = 0.$

A formalism for calculation of the two-loop effective action for the matrix model in arbitrary Cartan subalgebra-valued backgrounds will be established. As a  check of the formalism, we will evaluate the two-loop effective action in the case a specific background, namely that  appropriate to the physical situation in which a single D0-brane scatters from a stack of ($N$-1) D0-branes. In the fundamental representation, this corresponds to the choice
\be
g \,\vec{Y} =  {\rm diag} \left(\vec{r}_1(\t), \, \vec{r}_2(\t), \, \vec{r}_2(\t), \, \cdots, \, \vec{r}_2 (\t) \right),
\ee
where $\vec{r}_1(\t)$ denotes the position of the single D0-brane and $\vec{r}_2(\t)$ is the position of the stack of branes. Since the $U(1)$ degree of freedom associated with the centre-of-mass motion decouples, it suffices to consider the background
\be
g \, \vec{Y} =  {\rm diag} \left( \vec{r}_1(\t), \, \vec{r}_2 (\t), \, \vec{r}_2(\t), \, \cdots, \, \vec{r}_2(\t)  \right) - \vec{r}_{CM}(\t) \, {\bf 1_N}.
\ee
If the D0-brane is chosen to have unit mass, then the stack of branes has mass $N-1,$ so that
\be
g \, \vec{Y} =  \vec{r}(\t) \, H ,
\ee
where $\vec{r}(\t) = \vec{r}_1(\t) -  \vec{r}_2(\t)$ is the relative coordinate, and 
\be
H = \frac{1}{N} \, {\rm diag} \left(N-1,\, \, - \, {\bf 1_N} \right) = \sqrt{\frac{(N-1)}{N}} \, H_0.
\ee
This element $H$ of the Cartan subablgebra breaks $SU(N)$ to the subgroup $SU(N-1) \times U(1)$ generated by $\{H_{\Iu}, E_{\iu \ju}, H_0 \}.$ 
We can use $SO(9)$ invariance to parametrize the spatial separation of the branes as
\be
\vec{r}(\t) = (v \t, b, 0, \cdots, 0) ,
\ee
where  $b$ is the impact parameter for the scattering of the D0-brane from the stack   and $v$ is the relative speed. Note that the expression for the kernel (\ref{SUNkernel})   allows for the case of nonorthogonal  impact parameter and velocity; this is of significance in the case of three or more particle dynamics \cite{Helling:1999js}.

The Green's function (\ref{cGreen})  can be considered to be a matrix in the adjoint representation of $SU(N).$ Relative to the basis $(H_I, E_{0 \iu}, E_{\iu 0}, E_{\iu \ju}),$ the adjoint representation of the background field $\vec{Y}$ is
\be
g\vec{Y} = \vec{r}(\t) \, {\rm diag} \left( 0 \times {\bf 1_{N-1}}, \, {\bf 1_{N-1}}, \, - \, {\bf 1_{N-1}}, \, 0 \times {\bf 1_{(N-1)(N-2)}} \right).
\ee
As a result of this diagonal structure, the kernel (\ref{SUNkernel}), and thus the Green's function (\ref{cGreen}), are also diagonal as matrices in the adjoint representation  \cite{Kuzenko:2003wu},
\be
G^a{}_b = {\rm diag} \left(G^{(0)} {\bf 1_{N-1}}, \, G^{(1)} {\bf 1_{N-1}}, \, G^{(-1)} {\bf 1_{N-1}}, \, G^{(0)} {\bf 1_{(N-1)(N-2)}} \right),
\label{Gdecomp}
\ee
where the U(1) Green's function $G^{(e)}$ is defined by
\be 
G^{(e)} (\t, \t') = \ri \int_0^{\infty} \rd s \, K^{(e)}(\t,\t';s)
\label{Ge}
\ee
with 
\bea
K^{(e)}(\t,\t';s) &=&  \left( \frac{\ri}{4 \pi s}\right)^{\frac12} \, \left(\frac{2 e s v}{\sinh 2 e s v} \right)^{\frac12}  \re^{- \ri s e^2  b^2}  \nonumber \\ && \times \, \exp \left\{ - \frac{\ri e v}{2} (\t - \t')^2 \, \coth 2es v 
-  \ri e v \t \t' \tanh e s v \right\} \, I^{(e)}(\t,\t').
\label{Ke}
\eea
In (\ref{Gdecomp}), the values $e= 0, 1, -1$ and 0 which appear are the $U(1)$ weights of the basis vectors $(H_I, E_{0 \iu}, E_{\iu 0}, E_{\iu \ju})$ with respect to  $H,$  the element of the Lie algebra of $SU(N)$ to which the background is proportional.
The $U(1)$ Green's function $G^{(e)}$ satisfies the equation
\be
\left( - \, \frac{{\rm d}^2}{{\rm d}\t^2}  - e^2(b^2 + v^2 \t^2) \right) \, G^{(e)}(\t,\t') = - \delta(\t,\t').
\ee
Since the fields with $U(1)$ charge zero do not couple to the background, the propagator $G^{(0)}$ takes the simple form
\be
G^{(0)}(\t, \t') = \frac12 \, |\t - \t'|.
\label{G0}
\ee
There is no calculational advantage to using a proper time representation for  this massless free propagator. 

In the same background, the Green's function (\ref{aGreen}) takes the form 
\bea
G^{\m a}{}_{\n b} =  && i \int_0^{\infty} \rd s \,  {\rm diag} \left(   \delta^{\m}{}_{\n} \, K^{(0)} (s)\, {\bf 1_{N-1}}, \, (e^{-2  s  g F})^{\m}{}_{\n} \, K^{(1)} (s) \, {\bf 1_{N-1}}, \right. \nonumber \\
&& \left.  (e^{2  s g  F})^{\m}{}_{\n} \,K^{(-1)}(s) \,  {\bf 1_{N-1}},  \,
\delta^{\m}{}_{\n} \,K^{(0)} (s) \, {\bf 1_{(N-1)(N-2)}} \right) ,
\label{aG'}
\eea
with 
\be 
(e^{-2  s  g F})^{\m}{}_{\n} = \left( \begin{array}{ccc} \cosh 2s v  & \quad \sinh 2s v & \quad {\bf 0} \\
 \sinh 2s v & \quad \cosh 2s v  &  \quad {\bf 0} \\
{\bf 0} & \quad {\bf 0} & \quad {\bf 1_8} \end{array} \right),
\label{e^F}
\ee
and the the Green's function (\ref{psiGreen}) takes the form 
\bea
G^{a b}_{ \a \b}=  &&  i \int_0^{\infty} \rd s \, {\rm diag} \left(   (C^{-1})_{\a \b} \, K^{(0)}(s) \, {\bf 1_{N-1}}, \, (e^{\frac{s}{4}[\Gamma^{\m}, \Gamma^{\n}]  g F_{\m \n} }C^{-1})_{\a  \b} \, K^{(1)} (s) \,{\bf 1_{N-1}}, \right. \nonumber \\
&& \left.  (e^{- \frac{s}{4}[\Gamma^{\m}, \Gamma^{\n}] g  F_{\m \n} }C^{-1})_{\a \b} \,K^{(-1)} (s)\,  {\bf 1_{N-1}},  \, (C^{-1})_{\a \b} \,K^{(0)} (s)  \, {\bf 1_{(N-1)(N-2)}} \right) ,
\label{psiG'}
\eea
with 
\be 
(e^{\frac{s}{4}[\Gamma^{\m}, \, \Gamma^{\n}] g F_{\m \n} })_{\a}{}^{\b}  = \cosh  s v \,  \d_{\a}{}^{\b} + \sinh  s v \, (\Gamma^0 \Gamma^1)_{\a}{}^{\b} .
\ee

\sect{The structure of the two-loop graphs}
The two-loop contributions to the effective action come from Feynman diagrams which are of  ``figure-eight'' type or ``fish'' type. These diagrams arise from interaction terms in the action which are respectively quartic and cubic in the quantum fields, and are depicted in schematic form in Figure 1. 
\begin{figure}[!htb]
\begin{center}
\includegraphics{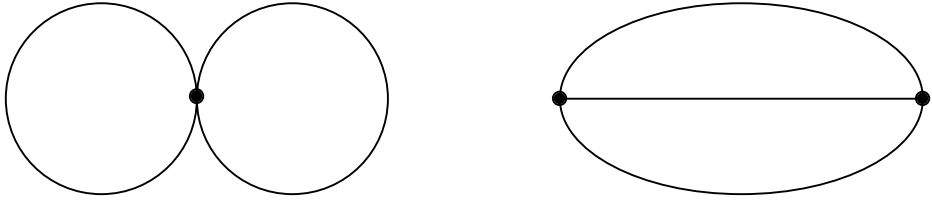}
\caption{``Figure-eight'' and ``fish'' type two-loop diagrams}
\end{center}
\end{figure}

There are no quartic vertices in the ghost action (\ref{Sg}), and the cubic vertices in it are 
\be
(S_{ghost})_3 = - \, g \, f_{a b d}  \, \int \rd \t \, \left(\bar{c}^{\, a} a^{\m b} (D_{\m}c)^d +\bar{c}^{\, a} (D^{\m} a_{\m})^b c^d \right),
\label{S3g}
\ee
where $f_{abc}$ are the structure constants of $SU(N),$ $[T_a, T_b] = \ri f_{ab}{}^c \, T_c.$
From the gauge-fixed action (\ref{Sgf}), the cubic interactions are
\be
(S + S_{g.f.})_3 = g \,   f_{abc} \, \int \rd \t \, \left( (D^{\m}a^{\n} )^a \,a_{\m}{}^{ b} a_{\n}{}^{ c} - \ri \psi^a (C \Gamma)^{\m} a_{\m}^b \psi^c \right)
\label{S3gf}
\ee
and the quartic interactions are 
\be
(S +S_{g.f.})_4 = - \,  \frac{g^2}{4} \, \int \rd \t \, f_{ab}{}^e \, f_{cde} \, a^{ a}_{\m} \, a^{ b}_{\n} a^{\m c} \, a^{\n d} .
\label{S4gf}
\ee

The figure-eight diagram arising from the quartic interaction (\ref{S4gf})  gives a contribution to the effective action of the form
\bea
\ri \, \Gamma_{I} = \frac{\ri g^2}{4} \int \rd \t \, \lim_{\t \rightarrow \t'} && \left\{ G^{\m a \, \n b}(\t, \t') \, \tr_{\rm A} \left( T_a T_b G_{\n \m}(\t, \t') \right) \right. \nonumber \\
- \, 2 && G^{\m a \, \n b}(\t, \t') \, \tr_{\rm A} \left( T_a T_b G_{\m \n}(\t, \t') \right) \nonumber \\
+ && \left. G^{\m a}{}_{\m}{}^{ b}(\t, \t') \, \tr_{\rm A} \left( T_a T_b G^{\n}{}_{\n}(\t, \t') \right) \right\} ,
\label{Sgfeight}
\eea
where the trace $\tr_{\rm A}$ is over gauge indices in the adjoint representation.

The fish diagram from the ghost sector leads to the following contribution to the effective action:
\bea
\ri \, \Gamma_{II} &=& - \, \frac{\ri  g^2}{2} \, \int \rd \t \int \rd \t' \, \left\{ G^{ \mu a \, \n b} (\t,\t') \,  {\rm tr}_{\rm A} \left( T_a \, D_{\m} G(\t,\t') \, T_b \, D\,'_{\n} G(\t',\t) \right) \right. \nonumber \\
 && + \, 2 \, D_{\m} G^{ \mu a \, \n b} (\t,\t') \,  \tr_{\rm A} \left( T_a \, G(\t,\t') \, T_b \, D\,'_{\n} G(\t',\t) \right) \nonumber  \\
&& + \,  \left. D_{\m} D\,'_{\n} G^{\mu a \, \n b} (\t,\t')\,  \tr_{\rm A} \left( T_a \, G(\t,\t') \, T_b \, G(\t',\t) \right) \right\} .
\label{Sgfish}
\eea

The fish diagrams involving involving scalar and vector propagators arising from the cubic interaction (\ref{S3gf})  give a contribution to the effective action 
\bea
\ri \, \Gamma_{III} & = & \frac{\ri g^2}{2} \, \int \rd \t \, G^{ \m a \, \n b } (\t,\t') 
  \left\{    \tr_{\rm A} \left( T_a \, D_{\m} D\,'_{\rho} G^{\s}{}_{\n}(\t,\t') \, T_b \, G^{\r }{}_{\s}(\t',\t) \right) \right. \nonumber \\ 
&& - \,  \tr_{\rm A} \left( T_a \, D_{\m} D\,'_{\n} G^{\r}{}_{ \s}(\t,\t') \, T_b \, G^{\s }{}_{\r}(\t',\t) \right) \nonumber \\
 && + \, \tr_{\rm A} \left( T_a \, D_{\s}  G_{\m}{}^{ \r}(\t,\t') \, T_b \, D\,'_{\r}G_{\n}{}^{ \s}(\t',\t) \right) \nonumber \\
 && - \, \tr_{\rm A} \left( T_a \, D_{\m}  G_{\s }{}^{\r}(\t,\t') \, T_b \, D\,'_{\r}G_{\n}{}^{ \s}(\t',\t) \right) \nonumber \\
&& - \,  \tr_{\rm A} \left( T_a \, D_{\s}  G_{\m}{}^{ \r}(\t,\t') \, T_b \, D\,'_{\n}G_{\r }{}^{\s}(\t',\t) \right) \nonumber \\
&& + \,  \left.  \tr_{\rm A} \left( T_a \, D_{\m}  G^{\s}{}_{ \r}(\t,\t') \, T_b \, D\,'_{\n}G^{\r}{}_{ \s}(\t',\t) \right) \right\} .
\label{Sgffish}
\eea

The fish diagram involving fermionic propagators arising from the cubic interaction (\ref{S3gf})  gives a contribution
\bea
\ri \, \Gamma_{IV} &=& \frac{\ri  g^2}{16} \, \int \rd \t \int \rd \t' \, (C \Gamma^{\m})^{\a \b} \,  (C \Gamma^{\n})^{\r \s} \,  G_{\mu \nu}^{a b}(\t, \t') \nonumber \\
&&  \times \, \tr_{\rm A} \left( T_a \, G_{\b \r} (\t, \t') \,T_b \, G_{\s \a} (\t', \t) \right).
\label{fermionicfish}
\eea

It should be noted at this point that no assumption has made about the background in the derivation of the above expressions for the two-loop contributions to the effective action. They are therefore suitable for use in analysis of scattering amplitudes for arbitrary configurations of $N$ D0-branes. In the remainder of the paper, we use the specific background discussed in the previous section, which is reflected in the manner in which the traces in the above expressions decompose into a number of separate $U(1)$ contributions. In particular, the fish diagrams will always contain one massless propagator. This is not  the case for more general choices of background, which will be treated in a separate publication.

The generic structure of the contributions to the effective action from the figure-eight diagram is
\be
\ri \, \Gamma = \int \rd \t \, \hat{G}^{ab}(\t,\t') \, \tr_{\rm A} \left( T_a  \, T_b \, \check{G}(\t, \t') \right),
\ee
where again any Lorentz structure has been suppressed, and  $\hat{G}$ and $\check{G}$ denote Green's functions or their derivatives.
As noted above, carrying out the trace decomposes this into a number of $U(1)$ components. For the background chosen in Section 3, the relevant group theoretic manipulations have been completed in earlier work on two loop graphs in superspace, see \cite{Kuzenko:2003wu}. The result is
\be
\ri \, \Gamma = N (N-1) \int \rd \t \, \lim_{\t \rightarrow \t'} \left\{  \hat{G}^{(e)}(\t,\t') \, \check{G}^{(e)}(\t, \t')  + \left( e \leftrightarrow -e \right) \right\} |_{e=1}.
\label{eightgroup}
\ee
Note that there are no massless propagators associated with the figure-eight diagram. Massless propagators would lead to divergences since the figure-eight diagram involves coincidence limits of propagators.

The generic structure of the contribution to the effective action from a fish diagram is
\be
\ri \, \Gamma = \int \rd \t \int \rd \t' \, G^{ab}(\t,\t') \, \tr_{\rm A} \left( T_a \, \hat{G}(\t,\t') \, T_b \, \check{G}(\t', \t) \right),
\ee
where any Lorentz structure has been suppressed, and  $\hat{G}$ and $\check{G}$ again denote Green's functions or their derivatives. Again making use of results from \cite{Kuzenko:2003wu}, these contributions decompose into $U(1)$ components in the form
\bea
\ri \, \Gamma = N (N-1) \int \rd \t \int \rd \t' \, \left\{  \right. & & G^{(0)}(\t,\t') \, \hat{G}^{(e)}(\t,\t') \, \check{G}^{(e)}(\t', \t)  \nonumber \\
+ && G^{(e)}(\t,\t') \, \hat{G}^{(-e)}(\t,\t') \, \check{G}^{(0)}(\t', \t)  \nonumber \\
+ && G^{(e)}(\t,\t') \, \hat{G}^{(0)}(\t,\t') \, \check{G}^{(e)}(\t', \t)  \nonumber \\
+  && \left. \left( e \leftrightarrow -e \right) \right\} |_{e=1}.
\label{fishgroup}
\eea

\sect{Evaluating the figure-eight diagram}
The two-loop contribution $\Gamma_I$ to the effective action from the figure-eight  diagram is given by (\ref{Sgfeight}).
Making use of the group theoretic results (\ref{eightgroup}) and substituting the propagator (\ref{aG'}),  
\bea
\ri \Gamma_{I}  &=& - \, N (N-1) \,\frac{\ri }{4} \int \rd \t \, \lim_{\t \rightarrow \t'}  \int_0^{\infty} \rd s \int_0^{\infty} \rd t \,  \left\{ \left[ \left( \re^{-2eg(s+t)F}\right)^{\m}{}_{\m}
 - 2 \left(\re^{2 eg (s-t) F} \right)^{\m}{}_{\m}  \right. \right. \nonumber \\
 && + \, \left. \left.  \left(\re^{-2 eg s F} \right)^{\m}{}_{\m} \, \left(\re^{- 2 eg t F} \right)^{\n}{}_{\n} \right]  K^{(e)}(\t, \t'; s) \, K^{(e)}(\t, \t'; t) + \left( e \leftrightarrow -e \right) \right\} |_{e=1},
\eea
where $s$ and $t$ are proper times associated with heat kernels.
From (\ref{Ke}), the coincidence limit of the heat kernel is  
\be
\lim_{\t \rightarrow \t'} K^{(1)}(\t, \t' ; s) = \left( \frac{\ri }{4 \pi s}\right)^{\frac12} \, \left(\frac{2  s v}{\sinh 2  s v} \right)^{\frac12} \, \re^{- \ri s  b^2} \,  \exp \left( - \ri  v \t^2 \tanh  sv \right).
\ee
For later comparison with existing two-loop calculations, we choose to evaluate the effective action in the Euclidean metric, which involves the Wick rotations
\be
\t \rightarrow \ri \t, \quad v \rightarrow - \ri v, \quad s \rightarrow - \ri s,  \quad t \rightarrow - \ri t, \quad \ri \Gamma_I \rightarrow - \Gamma_I.
\ee
The resulting expression for the (Euclidean) effective action is:
\bea
\Gamma_{I}  &=& N (N-1) \, \frac{\ri g^2 }{16 \pi }  \int \rd \t   \int_0^{\infty} \rd s \int_0^{\infty} \rd t \, 
\left( \frac{ - 4 v^2 }{ \sinh 2 s v \sinh 2 t v } \right)^{\frac12} \, \re^{- (s+t)  b^2 } \nonumber \\
&&\times \, \exp \left( -  v \t^2 (\tanh  s v + \tanh  t v) \right) \, \left\{ 4 \cosh 2(s+t) v - 8 \cosh 2(s-t) v  \right. \nonumber \\
&& \left. + \, 8 \cosh 2 s v \, \cosh 2 t v + 32 \cosh 2 s  v + 32 \cosh 2 t v \right\} .
\eea
The integral over (Euclidean) time $\t$ is Gaussian and so can be carried out in a straightforward manner. Also making use of the identity
\be
(\tanh  s v  + \tanh  t v ) \, \sinh 2  s v \, \sinh  2  t v = 4 \sinh  sv \,  \sinh  t v \, \sinh  (s+t) v,
\ee
we arrive at 
\bea
\Gamma_{I} & = & - \, N (N-1) \, \frac{  g^2 v^{\frac12 }}{4 \sqrt{\pi }}  \int_0^{\infty} \rd s \int_0^{\infty} \rd t \, \, \re^{-   (s+t)b^2} \nonumber \\
&& \times \, \left( \sinh s v \,  \sinh t v \, \sinh  (s+t) v \right)^{-\frac12}  
  \left\{ \cosh 2  (s+t) v - 2 \cosh 2  (s-t) v  \right. \nonumber \\
&& \left. + \, 2 \cosh 2  s v \, \cosh 2  t v + 8 \cosh 2  s v + 8 \cosh 2  t v   \right\} .
\eea
Introducing the new variables
\be 
x = \re^{2 s v} \quad {\rm and } \quad  y = \re^{2 t v},
\ee
the figure-eight contribution can be caste in the form
\bea
\Gamma_{I} & = &- \, N (N-1) \, \frac{g^2 }{8 \sqrt{2 \pi } v^{\frac32}}  \, \int_1^{\infty} \rd x \int_1^{\infty} \rd y \, 
 \frac{(x y) ^{-\frac{ b^2}{2 v} - \frac32} }{(x-1)^{\frac12} \, (y-1)^{\frac12}(xy-1)^{\frac12}}  \nonumber \\ 
&& \times \,\left\{  2 + 8 x + 8 y - x^2 - y^2  + 56 x y + 8 x^2 y + 8x y^2  + 2x^2y^2 \right\}.
\label{Gamma1}
\eea
As recognized in \cite{Becker:1998gp},  the remaining integrals can be performed explicitly and yield generalized hypergeometric functions of unit argument via the general result
\bea
&& \int_1^{\infty} \rd x \int_1^{\infty} \rd y \,
 \frac{(x-1)^{\a} \, (y-1)^{\b}}{(xy-1)^{\gamma}} \,  x^{-\zeta + \m} \, y^{-\zeta + \n} \nonumber \\
= && B(\a + 1, \b + 1) \, B(\zeta-\n-\b+\gamma -1, \a+\b-\gamma+2) \nonumber \\
&&   \times \, {}_3F_{2}(\a+1, \a + \b - \gamma +2, \a+\m-\n+1 ; \a + \b +2, \zeta+\a-\n+1 ; 1),
\label{hyper}
\eea
where $B(x,y) = \G(x) \G(y)/ \G(x+y).$
However, we find it more convenient to leave the result at this stage in the form (\ref{Gamma1}). 

It is important to note that the integrals in (\ref{Gamma1}) are nonsingular. There is no problem with convergence at the upper limits of the integrals since $\frac{b^2}{2 v} $ is positive; however, there is potentially a problem at the lower limits $x=1$ and $y=1.$ Since the numerator of the integrand in  (\ref{Gamma1}) is nonvanishing, it suffices to show that the integral
\be
\int_1^{\infty}  \rd x \int_1^{\infty} \rd y \, 
(x-1)^{-\frac12} \, (y-1)^{-\frac12} \, (xy-1)^{-\frac12}
\ee
is well-behaved at the lower limit. This can be achieved by the change of variable\footnote{We thank Sergei Kuzenko for assistance with this analysis.} 
$$\tilde{x}= (x-1)^{\frac12}, \quad \tilde{y}= (y-1)^{\frac12}, $$
which converts the integral to
$$ 4 \int_0^{\infty}  \rd \tilde{x} \int_0^{\infty} \rd \tilde{y} \, 
 ( \tilde{x}^2 \tilde{y}^2 +  \tilde{x}^2+ \tilde{y}^2)^{-\frac12}.$$
Introducing polar coordinates $r, \theta$ in the $\tilde{x}-\tilde{y}$ plane, the integral takes the form
$$ 4
 \int_0^{\infty}  \rd r \int_0^{\frac{\pi}{2}} \rd \theta\, 
 ( 1+r^2 \sin^2\theta \cos^2\theta)^{-\frac12},$$
which is clearly convergent. 

\section{Evaluating the Ghost Contribution}
The two-loop contribution $\Gamma_{II}$ to the effective action from the fish diagram involving ghost propagators is given in (\ref{Sgfish}). After integration of the third term by parts with respect to $D\,'_{\nu},$ the last two terms in (\ref{Sgfish}) become 
\bea
&-& \frac{\ri \, g^2}{2} \, \int \rd \t \int \rd \t'  \left\{  D_{\m} G^{ \mu a \, \n b} (\t,\t') \,  \tr_{\rm A} \left( T_a \, G(\t,\t') \, T_b \, D\,'_{\n} G(\t',\t) \right) \right. \nonumber  \\
&& - \, \left. D_{\m}  G^{\mu a \, \n b} (\t,\t')\,  \tr_{\rm A} \left( T_a \,D\,'_{\n} G(\t,\t') \, T_b \, G(\t',\t) \right) \right\} .
\eea
By transposing the argument of the trace and using the antisymmetry of the group generators in the adjoint representation, these two terms can be seen to cancel each other. Thus the surviving contribution to the effective action from the ghosts is
\be
\ri \, \Gamma_{II} = - \, \frac{\ri \, g^2}{2} \, \int \rd \t \int \rd \t' \,  G^{ \mu a \, \n b} (\t,\t') \,  {\rm tr}_{\rm A} \left( T_a \, D_{\m} G(\t,\t') \, T_b \, D\,'_{\n} G(\t',\t) \right) . 
\label{ghostgamma}
\ee
Using the group theoretic result (\ref{fishgroup}) and the Green's functions (\ref{Ge}),  (\ref{G0}) and  (\ref{aG'}), 
\bea
\ri \, \Gamma_{II} &=& - \, N (N-1) \,\frac{\ri g^2}{2} \,  \int \rd \t \int \rd \t'  \int_0^{\infty} \rd s \int_0^{\infty} \rd t  \, |\t - \t'| \, K^{(1)}(\t, \t'; s) \, K^{(1)}(\t', \t; t) \nonumber \\
& & \times \,\left\{ - v^2 \,\frac{[\t \cosh 2  s v - \t' ]}{\sinh 2  s v} \, \frac{[\t' \cosh 2  t v - \t ]}{\sinh 2  t v} +  v^2 \t \t' +  b^2 \right. \nonumber \\
&& + \, \left.  \ri \, \frac{(\t - \t')}{|\t - \t'|^2} \, \frac{2  v}{\sinh 2 t v} \left( \t \cosh 2  (s+t)v - \t' \cosh 2  s v \right) \right\} .
\eea
Substituting the expression (\ref{Ke}) for the $U(1)$ kernel, making the Wick rotations 
\be
\t \rightarrow \ri \t, \quad v \rightarrow - \ri v, \quad  s \rightarrow - \ri s, \quad t \rightarrow - \ri t, \quad \ri \, \Gamma_{II} \rightarrow - \, \Gamma_{II}
\label{Wick}
\ee
and  the changes of variables 
\be
x = \re^{2 s v}, \quad y =  \re^{2  t v}, \quad \t = \frac{ \xi + \r}{\sqrt{ v}}, \quad \t' = \frac{ \xi - \r}{\sqrt{ v}},
\label{changevariables}
\ee
\bea
\Gamma_{II} &=& - \, N (N-1) \,\frac{g^2}{2 \pi v^{\frac32}}  \, \int_{- \infty}^{\infty} \rd \xi \int_{- \infty}^{\infty} \rd \r \, \frac{|\r|}{\r} \int_1^{\infty} \rd x \int_1^{\infty} \rd y \,
\frac{(x y)^{-\frac{b^2}{2v} - \frac12}}{(x^2 - 1)^{\frac12} (y^2 -1)^{\frac12}} \, \nonumber \\
& & \times \,\exp \left( - \r^2 \left[\frac{y + 1}{y - 1} + \frac{x + 1}{x - 1} \right] - \xi^2 \left[\frac{y -1}{y+1} + \frac {x-1}{x+1} \right] \right) \nonumber \\
& &\times \, \left\{ - \r \, \frac{ b^2}{v} - \r \, (\xi^2 - \r^2) + \r \, \left( \xi \, \frac{x-1}{x+1} + \r \, \frac{x+1}{x-1} \right) \, \left( \xi \, \frac{y - 1}{y + 1} - \r \,  \frac{y+1}{y-1} \right) \right. \nonumber \\
& & + \left. \frac{y}{y^2-1} \left( \xi \, (xy + \frac{1}{xy} -x -\frac{1}{x}) + \r \, (xy + \frac{1}{xy} +x +\frac{1}{x}) \right) \right\}.
\eea
We  gave also made use of the fact that the parallel displacement propagator has the property $I^{(1)}(\t, \t') \, I^{(1)}(\t', \t) = 1$   \cite{Kuzenko:2003eb}.
The integrals over $\xi$ and $\r$ are Gaussian and so are straightforward to carry out. The result is 
\bea
\Gamma_{II} &=&  N (N-1) \, \frac{g^2}{8 \sqrt{2 \p} v^{\frac32}} \, \int_{1}^{\infty} \rd x \int_{1}^{\infty} \rd y \,  \frac{(x - 1)^{\frac12} (y -1)^{\frac12}}{(xy-1)^{\frac32}} \, (x y)^{-\frac{b^2}{2v} - \frac32} \nonumber \\ 
& & \times \, \left\{2\,  \frac{ b^2}{v} xy +3 \, \frac{(x^2y+xy^2)}{(xy-1)} -2 \, \frac{(x^2y^2 + y)}{(y-1)} \right\}.
\label{GammaII}
\eea
The remaining integrals can be done using the general result (\ref{hyper}), but again we choose to leave the result expressed in the above form (\ref{GammaII})

As in the previous section, it is possible to check that the contribution (\ref{GammaII}) to the effective action is free from divergences. However, rather than detailing the checks for each particular diagram, an explicit check will only be demonstrated for the total bosonic and ghost contribution collated in section 8, and for the fermionic contribution computed in section 9. 

\sect{Evaluating the fish diagrams from the scalar sector}
The two-loop fish diagrams arising from the cubic interaction involving scalars and vectors give the contribution $\Gamma_{III}$ in (\ref{Sgffish}) to the effective action. Using the cyclic property of the trace and making the substitutions $\mu \leftrightarrow \nu$ and $ \t \leftrightarrow \t',$ the the fifth term in (\ref{Sgffish}) can be caste in the same form as the fourth. In order to avoid the necessity to calculate second derivatives of heat kernels, it is convenient to integrate the first term in (\ref{Sgffish}) by parts. This results in two terms, one of which can also be caste in the form of the fourth term of (\ref{Sgffish}) by cycling the trace and relabeling of variables. Integrating the second term in (\ref{Sgffish}) by parts also yields two terms, one of which can be re-caste in the form of the last term in (\ref{Sgffish}). The result is that the contribution to the effective action is 
\bea
\ri \, \Gamma_{III} & = & \frac{\ri \, g^2 }{2} \, \int \rd \t \, G^{ \m a \, \n b } (\t,\t') 
\left\{   - \,   \tr_{\rm A} \left( T_a \, D_{\m}  G^{\s}{}_{ \n}(\t,\t') \, T_b \, D\,'_{\rho} G^{\r }{}_{\s}(\t',\t) \right) \right. \nonumber \\ 
&  &  + \, \tr_{\rm A}  \left( T_a \, D_{\s}  G_{\m \n}(\t,\t') \, T_b \, D\,'_{\r} G^{\r \s}(\t',\t) \right) \nonumber \\
 & & + \, \tr_{\rm A} \left( T_a \, D_{\s}  G_{\m}{}^{ \r}(\t,\t') \, T_b \, D\,'_{\r}G_{\n}{}^{ \s}(\t',\t) \right) \nonumber \\
 & & - \,3 \, \tr_{\rm A} \left( T_a \, D_{\m}  G_{\s }{}^{\r}(\t,\t') \, T_b \, D\,'_{\r}G_{\n}{}^{ \s}(\t',\t) \right) \nonumber \\
& &  + \, 2 \left.  \tr_{\rm A} \left( T_a \, D_{\m}  G^{\s}{}_{ \r}(\t,\t') \, T_b \, D\,'_{\n}G^{\r }{}_{\s}(\t',\t) \right) \right\} .
\label{GammaIII0}
\eea
Under certain assumptions about the nature of the background, the first and fourth terms in (\ref{GammaIII0}) are proportional to each other, despite the fact that they have different Lorentz structure. The  proof is provided in Appendix B, and applies provided the background is of the form $g \,A_{\m}(\t) = (0, \vec{v} \t + \vec{b} ) \, H,$ where $H$ is an arbitrary element of  the Cartan subalgebra. The resulting simplification of the structure of  (\ref{GammaIII0}) therefore applies for arbitrary configurations of $N$ D0-branes, and not just the special case considered in detail in this paper. The first term in (\ref{GammaIII0}), which has a coefficient $-1$, combines with the fourth term, which has a coefficient $-3$, to yield a single term with coefficient $-4$:  

\bea
\ri \, \Gamma_{III} & = & \frac{\ri  g^2 }{2} \, \int \rd \t \, G^{ \m a \, \n b } (\t,\t') 
\left\{   - \,   4 \,\tr_{\rm A} \left( T_a \, D_{\m}  G^{\s}{}_{ \n}(\t,\t') \, T_b \, D\,'_{\rho} G^{\r }{}_{\s}(\t',\t) \right) \right. \nonumber \\ 
&  &  + \, \tr_{\rm A} \left( T_a \, D_{\s}  G_{\m \n}(\t,\t') \, T_b \, D\,'_{\r} G^{\r \s}(\t',\t) \right) \nonumber \\
 & & + \, \tr_{\rm A} \left( T_a \, D_{\s}  G_{\m}{}^{ \r}(\t,\t') \, T_b \, D\,'_{\r}G_{\n}{}^{ \s}(\t',\t) \right) \nonumber \\
& &  + \, 2 \left.  \tr_{\rm A} \left( T_a \, D_{\m}  G^{\s }{}_{\r}(\t,\t') \, T_b \, D\,'_{\n}G^{\r}{}_{\s}(\t',\t) \right) \right\} 
 \nonumber \\
& & \equiv \ri \, \Gamma_{III}^{(1)} + \ri \, \Gamma_{III}^{(2)} + \ri \, \Gamma_{III}^{(3)} +
 \ri \, \Gamma_{III}^{(4)}.
\label{GammaIII1}
\eea

Using the group theoretic result (\ref{fishgroup}), each of these five contributions can be expressed in terms of $U(1)$ Green's functions. For the purposes of calculation, we choose to divide each of these contributions into two pieces, depending upon whether or not the undifferentiated Green's function carries a nonzero $U(1)$ charge. In the case of $ \Gamma_{III}^{(1)},$ these two contributions will be denoted $  \Gamma_{III}^{(1)}(1)$ and $\Gamma_{III}^{(1)}(0), $ with the same notation for the other four terms in (\ref{GammaIII1}). For the purposes of brevity, we also introduce the notation
\bea
X &=& - \, N (N-1) \, \frac{\ri \, g^2}{2}\,  \int \rd \t \int \rd \t'  \int_0^{\infty} \rd s \int_0^{\infty} \rd t  \, |\t - \t'| \, K^{(1)}(\t, \t'; s) \, K^{(1)}(\t', \t; t) \nonumber \\
I &=&  v^2 \, X \, \left\{ \frac{[\t \cosh 2  s v - \t']}{\sinh 2  s v} \,  \frac{[\t' \cosh 2  t v - \t]}{\sinh 2  t v} - \t \t' \right\} \nonumber \\
J &=&  v^2 \, X \, \t' \, \frac{[\t \cosh 2  s v - \t']}{\sinh 2 s v} \,  \nonumber \\
K &=& \ri  v \, X \, \frac{(\t - \t')}{|\t - \t'|^2}\, \frac{[\t \cosh 2  t v - \t' ]}{\sinh 2  t v}  \nonumber \\
L &=&  \ri v \, X \, \t \, \frac{(\t - \t')}{|\t - \t'|^2} \nonumber \\
M &=& b^2 \, X. 
\label{notation}
\eea
The various contributions to the effective action are then\footnote{In this notation, the two-loop ghost contribution (\ref{ghostgamma}) is $i \, \Gamma_{II} = 2 K \cosh 2tv + 2 L \sinh 2 t v -I+M.$}:
\bea
\ri \, \Gamma_{III}^{(1)}(0) &=& - 4\, I \, \cosh2(s+t) v + 4 M \nonumber \\
\ri \, \Gamma_{III}^{(1)}(1) &=& 4 K \,( \cosh2(s+t) v +  \cosh2(s-t) v) + 4 L \, (\sinh2(s+t) v 
+   \sinh2(s-t) v) \nonumber \\
\ri \, \Gamma_{III}^{(2)}(0) &=& (I \, \cosh 2  t v - J \,\sinh 2  t v - M) (2 \cosh2  s v + 8)   + J \, \sinh 2  sv \, (2 \cosh2  t v + 8)    \nonumber \\
\ri \, \Gamma_{III}^{(2)}(1) &=& - K \, (2 \cosh2  (s+t) v + 8) + (- K \, \cosh 2  t v +  L \,  \sinh 2  t v ) \, (2 \cosh2  s v + 8)   \nonumber \\
\ri \, \Gamma_{III}^{(3)}(0) &=& I \, \cosh 2  (s-t) v +2 \, J \, \sinh 2  (s-t) v - M \nonumber \\
\ri \, \Gamma_{III}^{(3)}(1) &=& - 2 \, K \, \cosh 2  (s+t) v + 2 \, L \, \sinh 2  (s+t) v \nonumber \\
\ri \, \Gamma_{III}^{(4)}(0) &=& 2 \, (I - M)\, (2 \cosh 2  (s+t) v + 8) \nonumber \\
\ri \, \Gamma_{III}^{(4)}(1) &=& -4\, (K \, \cosh 2  s v + L \, \sinh 2  s v)\, (2 \cosh 2 t v + 8).
\eea 
After the Wick rotation (\ref{Wick}) and the change of variables (\ref{changevariables}), these expressions can be expressed as integrals which can be evaluated in terms of generalized hypergeometrics. With the notation
\be
Q = N (N-1) \,  \frac{g^2}{8 \sqrt{2 \p } v^{\frac32}} \, \int_{1}^{\infty} \rd x \int_{1}^{\infty} \rd y \,  \frac{(x - 1)^{\frac12} (y -1)^{\frac12}}{(xy-1)^{\frac32}} \, (x y)^{-\frac{b^2}{2v} - \frac32} \quad ,
\label{Q}
\ee
the integrals are (after making use of the symmetry $x \leftrightarrow y$ in $Q$ to combine some terms)
\bea
 \Gamma_{III}^{(1)}(0) &=&   12 \, Q \, (xy-1)^{-1} (x^3y^2+ y) + 8 \frac{ b^2}{v} Q\, x y
\nonumber \\
\Gamma_{III}^{(1)}(1) & = &- \, 2 \, Q \, (y-1)^{-1} (x^2y^3 + x^2y + y^3  +  y + x^2y^2 + x^2  + y^2 + 1)   -  2\, Q \, (x^2y^2 - 1)  \nonumber \\
\Gamma_{III}^{(2)}(0) &=& - \, Q \, (xy-1)^{-1} (3 x^3 y^2 + 6x^2y + 3 y +24 x^2y^2 + 24xy) - 2 \frac{b^2}{v} Q \, (x^2y + y +8xy) \nonumber \\
 \Gamma_{III}^{(2)}(1) &=& \frac12 \, Q \,  (y-1)^{-1} (3 x^2 y^3 + 3 y+ 8 x  +x^2y +24 xy^2 + y^3 + 8 xy^3+24xy + 3x^2 y^2 + 3  \nonumber \\  & & - \, x^2  +  y^2) - \frac12 \, Q \, (x^2y^2 - 1+ 8 xy^2 - 8 x) \nonumber \\
\Gamma_{III}^{(3)}(0) &=& - \,6 \, Q \,  (xy-1)^{-1} x^2y - 2 \frac{b^2}{v} Q \, xy\nonumber \\
 \Gamma_{III}^{(3)}(1) &=&  \, Q \,  (y-1)^{-1} (x^2 y^3 + y + x^2 y^2  + 1) - Q \, (x^2y^2 - 1)   \nonumber \\
\Gamma_{III}^{(4)}(0) &=& -12 \, Q \,  (xy-1)^{-1} (x^3 y^2 + y + 8 x^2y) - 4  \frac{b^2}{v} Q \, (x^2y^2 + 1 +8xy) \nonumber \\
\Gamma_{III}^{(4)}(1) &=& 2 \, Q \,  (y-1)^{-1} (x^2y^3 + 9x^2y + 9 x^2 y^2 + y^3   +  9 y+ x^2  + 9 y^2  + 1 )  \nonumber \\  & & + \, 2 \, Q \, ( x^2y^2 - 1 + 8 x^2y - 8 y)   .
\label{GammaIII}
\eea 

\sect{The total bosonic and ghost contribution}
Combining the results (\ref{Gamma1}),  (\ref{GammaII}) and (\ref{GammaIII}) for $\Gamma_I,$ $\Gamma_{II}$ and $\Gamma_{III},$ the total contribution to the effective action from two-loop diagrams involving only bosonic and ghost quantum fields is:
\bea
\Gamma_B &=& - \, Q \, \left\{ \frac32 x^2y^2 - 12x^2y + 12y - \frac32 \right\} \nonumber \\
&& + \,\frac{Q}{(y-1)} \, \left\{  \frac52 x^2y^3 + 4 xy^2+\frac{33}{2} x^2y^2+\frac12 y^3+\frac{33}{2} x^2 y + 12 xy^2 \right. \nonumber \\
&& + \, \frac12 x^2 +\frac{33}{2} y^2+4x 
+ \left. \frac{33}{2}y + 12xy+\frac52 \right\} \nonumber \\
&& - \, \frac{Q}{(xy-1)} \, \left\{3x^3y^2+24x^2y^2+102x^2y+24xy+3y \right\} \nonumber \\
&& - \, \frac{Q\, (xy-1)}{(x-1)(y-1)}\, \left\{ 2x^2y^2+16x^2y -2x^2 +56xy+16y+2 \right\} \nonumber \\
&& - \,\frac{b^2}{v} \, Q \, \left\{4x^2y^2+2x^2y +40xy +2y+4 \right\} .
\eea

It is convenient to combine all of these terms. The first step is to eliminate the explicit $\frac{b^2}{v}$ dependence in the last term by making use of the identity
\bea
 \z \, h(\a, \b,\g,\m,\n) &= &\b \, h(\a, \b -1,\g,\m,\n+1) +(\n+1) \, h(\a, \b,\g,\m,\n)  \nonumber \\
&-&  \g\, h(\a, \b,\g+1,\m+1,\n+1 )
\label{ident1}
\eea
with 
\be 
h(\a, \b,\g,\m,\n) = \int_1^{\infty} \rd x \int_1^{\infty} \rd y \,
 \frac{(x-1)^{\a} \, (y-1)^{\b}}{(xy-1)^{\gamma}} \,  x^{-\zeta + \m} \, y^{-\zeta + \n}
\ee
and 
\be 
\z = \frac{b^2}{2v}.
\ee
This is easily proven by integration by parts with respect to $y.$ Next, all terms are placed on a common denominator. The result is
\bea
\Gamma_B &=& - \, \frac{Q}{(x-1)(y-1)(xy-1)} \, \left\{ 5 (1+x^4y^4) + 38xy(1+x^2y^2) - 20 y(1+x^3y^3) \right.  \nonumber \\ &  & + \, \left. 212 x^2 y (1+xy) - 2 x^2 (1+x^2y^2) -92x^3 y - 374 x^2y^2\right\}.
\label{GammaB}
\eea

Although positive powers  of $(x-1),$ $(y-1)$ and $(xy-1)$ in the denominator mean that this expression is potentially divergent at $x=1$ and $y=1,$ it it is in fact not so. This is again easily seen with the change of variable $(x-1)^{\frac12} = r \, \cos \theta, (y-1)^{\frac12} = r \, \sin \theta.$ Then
\bea
 \Gamma_B & = & - \, N (N-1) \, \frac{g^2}{2 \sqrt{2 \pi} v^{\frac32}}  \int_0^{\infty} \rd r \, r \int_0^{2 \pi} \, \rd \theta \, r^{-5} (1 + O(r^2))  \, \left\{ 136 r^2 \cos^2 \theta - 136 r^2 \sin^2 \theta \right. \nonumber \\  
 && + \,\left. 268 r^4 \cos^4 \theta - 140 r^4 \sin^4 \theta + 128 r^4 \cos^2 \theta \sin^2 \theta + O(r^6) \right\}.
\eea
The terms of order $r^2$ in the curly brackets cancel under the substitution $y \leftrightarrow x$ (or $\cos^2 \theta \leftrightarrow \sin^2 \theta$) in the second one, so the curly brackets are proportional to $r^4,$ and $\Gamma_B$ is nonsingular at the lower limit $r \rightarrow 0$ of the $r$ integral. For later use, the leading term is
\be
\Gamma_B  =  - \, 128 N (N-1) \, \frac{g^2}{2 \sqrt{2 \pi} v^{\frac32}}  \int_0^{\infty} \rd r \,  \int_0^{2 \pi} \rd \theta  \, \left\{   \cos^4 \theta + \cos^2 \theta \sin^2 \theta + O(r^2) \right\}.
\label{Bleading}
\ee

\sect{Evaluating the  fermionic diagrams}
The contribution $\Gamma_{IV}$ to the two-loop effective action from diagrams involving fermionic propagators is given by (\ref{fermionicfish}). Using the group theoretic result (\ref{fishgroup}), this naturally splits into two pieces: $ \Gamma_{IV}(0),$ in which the bosonic propagator in the diagram has $U(1)$ charge zero, and $\Gamma_{IV}(1),$ in which the bosonic propagator in the diagram carries $U(1)$ charge $e=1.$ Substituting the fermionic propagator (\ref{psiG'}), and with ``$\tr$'' denoting a trace over spinor indices
\bea
\ri \, \Gamma_{IV}(0) &=& N (N-1) \, \frac{\ri g^2 }{4}\,   \int \rd \t \int \rd \t'  \int_0^{\infty} \rd s\int_0^{\infty} \rd t  \, \, |\t - \t'| \, K^{(1)}(\t, \t'; s) \, K^{(1)}(\t', \t; t ) \nonumber \\
& & \times \, \tr \left\{ \Gamma^{\m} \, \left( - \ri  v \, \Gamma^0 \,  \frac{[\t \cosh 2  s v - \t' ]}{\sinh 2  s v}  + \ri  v \t \, \Gamma^1 + \ri  b \, \Gamma^2 \right)   \right. \nonumber \\
&& \times \, \left({\bf 1_{16}} \cosh  s v + \Gamma^0 \Gamma^1 \, \sinh  s v \right) \, \Gamma_{\m} \,  \left( - \ri  v \, \Gamma^0 \,  \frac{[\t' \cosh 2  t v - \t ]}{\sinh 2  t v}  + \ri  v \t' \, \Gamma^1 + \ri  b \, \Gamma^2 \right) \nonumber \\
&&\times \, \left.  \left({\bf 1_{16}} \cosh  t v + \Gamma^0 \Gamma^1 \, \sinh  t v \right) \right\},
\eea
and
\bea
\ri \, \Gamma_{IV}(1) &=& - \, N (N-1)  \, \frac{ g^2 }{4} \,   \int \rd \t \int \rd \t'  \int_0^{\infty} \rd s \int_0^{\infty} \rd t  \, \frac{(\t - \t')}{|\t - \t'|} \, K^{(1)}(\t, \t'; s) \, K^{(1)}(\t', \t; t)  \, \nonumber \\
& &\times \, \left( \re^{-2gsF}\right)^{\m}{}_{\n} \, \tr \left\{ \Gamma_{\m} \, \left(  v \, \Gamma^0 \,  \frac{[\t \cosh 2  t v - \t' ]}{\sinh 2  t v}  +   v \t \, \Gamma^1 +  b \, \Gamma^2 \right)   \right. \nonumber \\
&& \times \, \left({\bf 1_{16}} \cosh  t v - \Gamma^0 \Gamma^1 \, \sinh  t v \right) \, \Gamma^{\n} \Gamma^0 \nonumber \\
&& - \, \Gamma_{\m} \Gamma^0 \Gamma^{\n}  \, \left(  v \, \Gamma^0 \,  \frac{[\t' \cosh 2  t v - \t]}{\sinh 2  t v}  -   v \t' \, \Gamma^1 -  b \, \Gamma^2 \right) \,  \nonumber \\
&& \times \,  \left. \left({\bf 1_{16}} \cosh  t v +\Gamma^0 \Gamma^1 \, \sinh  t v \right) \,  \right\}.
\eea
Carrying out the traces over spinor indices results in the following expressions (using the notation (\ref{notation})):
\bea
\ri \, \Gamma_{IV}(0) &=&  \, \tr ({\bf 1_{16}}) \, \left\{ 4 \, (I - M) \, \cosh (s-t) v + 8 \, J \, \sinh  (s-t)v \right. \nonumber \\
&  &- \, \left.  6 \, M \sinh s v \, \sinh  t v \right\}
\eea 
and 
\be
\ri \, \Gamma_{IV}(1) =   \tr ({\bf 1_{16}}) \, \left\{ -8 \, K \cosh e t v +8 \, L \sinh e t v  \right\}. \quad \quad \quad \quad \quad
\ee 
Again, after the Wick rotation (\ref{Wick}) and the change of variables (\ref{changevariables}), these expressions yield integrals which can be evaluated in terms of hypergeometric functions. With the notation (\ref{Q}), 
\bea
\Gamma_{IV}(0) & = &  \tr ({\bf 1_{16}}) \, Q \, (xy -1)^{-1}(2 x^{\frac32} y^{\frac12} - 2 x^{\frac12}y^{\frac32} - 24 x^{\frac32}y^{\frac32} - 2 x^{\frac52}y^{\frac32} +2  x^{\frac32}y^{\frac52})   \quad \quad \quad \quad \quad \quad \quad  \nonumber \\
&& + \,  \tr ({\bf 1_{16}})\, \frac{b^2}{v} \, Q \, (3 x^{\frac12} y^{\frac12} -  x^{\frac32}y^{\frac12} -  x^{\frac12}y^{\frac32} + 3 x^{\frac32}y^{\frac32})
\eea 
and 
\be
 \Gamma_{IV}(1) =  \tr ({\bf 1_{16}}) \, Q \, (y -1)^{-1} (4 xy^{\frac52} + 8 xy^{\frac32} + 4 xy^{\frac12})  - \, \tr ({\bf 1_{16}}) \, Q \,  (4 xy^{\frac32} - 4 xy^{\frac12}).
\ee 

Collating these results and expressing them on a common denominator by use of the identity (\ref{ident1}), the total contribution to the two-loop effective action from diagrams involving fermionic quantum fields is
\bea
\Gamma_F & = &  \tr ({\bf 1_{16}}) \, Q \, \frac{x^{\frac12} y^{\frac12}}{(x-1)(y-1)(xy-1)} \, \left\{ -x(1+x^2y^2) +16xy^{\frac12}(1+x^{\frac32}y^{\frac32}) \right. \nonumber \\
&& - \left. 9xy(1+xy) -16xy^{\frac32}(1+x^{\frac12}y^{\frac12}) -3x^2(1+xy)+26x^2y\right\}.
\label{GammaF}
\eea

Again, positive powers  of $(x-1),$ $(y-1)$ and $(xy-1)$ allow for a potential divergences at $x=1$ and $y = 1.$  The absence of such a divergence easily seen with the same change of variable used previously, $(x-1)^{\frac12} = r \, \cos \theta, (y-1)^{\frac12} = r \, \sin \theta.$ Then
\bea
 \Gamma_F & = & 16 N (N-1) \, \frac{g^2}{2 \sqrt{2 \pi} v^{\frac32}}  \int_0^{\infty} \rd r \, r \int_0^{2 \pi} \rd \theta \, r^{-5} (1 + O(r^2))  \, \left\{ 6 r^2 \cos^2 \theta -  6r^2 \sin^2 \theta \right. \nonumber \\  
&& + \, \left. 16 r^2 \cos^2  \theta \sqrt{1+ r^2 \cos^2 \theta} - 16 r^2 \sin^2  \theta \sqrt{1+ r^2 \sin^2 \theta}
 \right. \nonumber \\  
 && + \, \left. 18 r^4 \cos^4 \theta - 10 r^4 \sin^4 \theta + 8 r^4 \cos^2 \theta \sin^2 \theta + O(r^6) \right\}.
\eea
The terms of order $r^2$ in the curly brackets cancel under the substitution $y \leftrightarrow x$ (or $\cos^2 \theta \leftrightarrow \sin^2 \theta$) in the second and fourth term, so the curly brackets are proportional to $r^4,$ and $\Gamma_F$ is nonsingular at the lower limit $r \rightarrow 0$ of the $r$ integral. In fact, the leading term is 
\be
\Gamma_F = 128 N (N-1) \, \frac{g^2}{2 \sqrt{2 \pi} v^{\frac32}}  \int_0^{\infty} \rd r \,  \int_0^{2 \pi} \rd \theta  \, \left\{   \cos^4 \theta + \cos^2 \theta \sin^2 \theta  \right\}.
\ee
which precisely cancels the leading  bosonic contribution (\ref{Bleading}) to the effective action.

\sect{Comparison with the results of Becker and Becker}
In \cite{Becker:1998gp}, the two-loop effective action for the matrix model was computed  for gauge group  $SU(2).$ Here, we show that the results of this paper for $SU(N)$ reproduced the Becker's result in the case $N=2.$ There is a subtlety in the comparison of the results, which relates to the normalization of the generators in the fundamental of $SU(2).$ The normalization used in \cite{Becker:1998gp}  is $ {\rm tr_F} (T_a T_b) = \frac12 \delta_{ab},$ with structure constants determined by  $[T_a, T_b] = \ri \, \e_{ab}{}^c \, T_c,$ while the conventions in the present paper are $ {\rm tr_F} (T_a T_b) = \delta_{ab},$ with  $[T_a, T_b] = \ri \,\sqrt{2} \, \e_{ab}{}^c \, T_c.$ This is equivalent to a rescaling of the coupling constant by a factor of $\sqrt{2},$ since in the notation in  \cite{Becker:1998gp}, $(D_{\m} \phi^a) \,T_a = (\partial_{\m} \phi^a + \ri g \e_{bc}{}^a \,A_{\m}^b \,\phi^c) \, T_a,$ while in this paper, $(D_{\m} \phi^a) \,T_a = (\partial_{\m} \phi^a + \ri g \sqrt{2} \, \e_{bc}{}^a \,A_{\m}^b \, \phi^c) \, T_a.$ The two-loop  contributions to the effective action are proportional to $g^2,$ which means that the results in this paper must be divided by a factor of 2 for comparison with the results in \cite{Becker:1998gp}.

The contribution to the two-loop effective action for the $SU(2)$ matrix model from bosons and ghosts in \cite{Becker:1998gp} comes from the terms odd in $v$ in the expression
\bea
\Gamma_B &=& \frac{ \sqrt{2} \pi}{3 v^{3/2}} \,  \frac{\Gamma(\z -\frac12 )}{\Gamma(\zeta)} \, \left\{ \frac{49}{8} \,{}_3F_2(\frac12, \frac12, \frac12; 1, \zeta;1) - {}_3F_2(\frac32, \frac32, \frac32; 3, \zeta + 1;1) \right. \nonumber \\
&& - \, \left.\frac{137}{16} \, \frac{(2\zeta-1)}{\z }\,   {}_3F_2(\frac12, \frac12, \frac12; 1, \zeta+1 ;1) \right\},
\eea
where
\be
\zeta = \frac{b^2}{2v}.
\ee
Without the need to select the terms odd in $v$ by hand, the result can be expressed 
\bea
\Gamma_B &=& \frac{ \sqrt{2} \pi}{3 v^{3/2}} \,  \frac{\Gamma(\zeta -\frac12)}{\Gamma(\zeta)} \, \left\{ \frac{49}{16}  \, {}_3F_2(\frac12, \frac12, \frac12; 1, \zeta;1) +  \frac{49}{16}\, \frac{(\zeta+\frac12)\zeta-\frac12)}{\zeta(\zeta+1)} \, {}_3F_2(\frac12, \frac12, \frac12; 1, \zeta+2;1) \right. \nonumber \\
&& - \, \frac12 \, {}_3F_2(\frac32, \frac32, \frac32; 3, \zeta + 1 ;1) 
- \frac18 \frac{(2 \zeta - 1)(2 \zeta + 1)}{(\zeta+1)(\zeta+2)} \, {}_3F_2(\frac32, \frac32, \frac32; 3, \zeta+3;1) 
\nonumber \\
&& - \,\left.\frac{137}{16} \, \frac{(2\zeta-1)}{\zeta}\,   {}_3F_2(\frac12, \frac12, \frac12; 1, \zeta + 1;1) \right\}.
\label{BBB}
\eea
The hypergeometics  can be written in terms of double integrals using (\ref{hyper}): 
\be
\Gamma_B = \frac{ Q}{3}  \, \left\{  \frac{(xy-1)}{(x-1)(y-1)} \left(\frac{49}{2}(1+x^2 y^2 ) - 137 x y \right)- 64 \, \zeta \, (1+x^2 y^2) \right\},
\ee
where $Q$ is defined by (\ref{Q})  (with $N$=2 in this case).

Eliminating the explicit $\zeta$ terms using (\ref{ident1}) and placing all the terms on a  common denominator, 
\bea
\Gamma_B  &=& \frac{Q}{(x-1) (y-1) (xy-1)} \, \left\{- \frac{5}{2} (1 + x^4y^4) + \frac{32}{3} x (1+x^3y^3)  - \, \frac{58}{3} x y (1+x^2y^2) \right. \nonumber \\
&&- \, \left. \frac{224}{3} x^2y(1+xy) + \frac{515}{3} x^2y^2 \right\}.
\label{Beckerbosonic}
\eea

Although not immediately obvious, this is equivalent to the $SU(N)$ result (\ref{GammaB}) for the case $N$=2 (after the rescaling discussed at the start of this section). Establishing the equivalence requires the use of a set of identities which we now derive. The result (\ref{ident1}) was derived by integration by parts with respect to $y.$ Integrating by parts with respect to $x$ gives rise to the related identities
\bea
 \z \, h(\a, \b,\g,\m,\n) &=& \a \, h(\a-1, \b ,\g,\m+1,\n) +(\m+1) \, h(\a, \b,\g,\m,\n) \nonumber \\
&& - \, \g\, h(\a, \b,\g+1,\m+1,\n+1 ).
\label{ident2}
\eea
Subtracting (\ref{ident1}) from (\ref{ident2}) gives the new identities
\bea
0 &=&  \a \, h(\a-1, \b ,\g,\m+1,\n) - \b \, h(\a, \b-1 ,\g,\m,\n+1) + (\m-\n) \, h(\a, \b,\g,\m,\n) \nonumber \\ &\equiv & f(\a, \b,\g,\m,\n).
\label{ident3}
\eea
The difference between the $SU(N)$ result (\ref{GammaB}) (appropriately rescaled) and the Becker's result (\ref{Beckerbosonic}) can be expressed as 
\bea
\frac{g^2}{8 \sqrt{2 \pi} v^{\frac32}}  && \left[ -\, \frac43 \, f( \frac12, \frac12, \frac52, -\frac12, -\frac32) + \frac{184}{3} \, f( \frac12, \frac12, \frac52, \frac12, -\frac12) \right. \nonumber \\
&& - \, \left. \frac43 \, f( \frac12, \frac12, \frac52, \frac32, \frac12) \right],
\eea
which vanishes identically by (\ref{ident3}).

The contribution from  the fermions as calculated in \cite{Becker:1998gp} is 
\bea
\Gamma_F &=&- \,  \frac{4 \sqrt{2} \pi}{v^{3/2}} \,  \frac{\Gamma(\zeta+1 )}{\Gamma(\z + \frac32 )} \, {}_3F_2(\frac12, \frac32, \frac32; 3, \z + \frac32 ;1) \nonumber \\ 
&& -  \frac{135 \sqrt{2} \pi}{512 \, v^{3/2}} \, \frac{\Gamma( \zeta + 1)}{\Gamma(\z + \frac72  )} \, {}_3F_2(\frac52, \frac52, \frac72; 5, \z + \frac72;1) \nonumber \\
&& + \, \frac{32 \sqrt{2} \pi}{v^{3/2}} \, \frac{(- \Gamma(\z + \frac12  )^2 + \Gamma(\zeta )\Gamma(\zeta +1))}{\Gamma(\zeta )\Gamma(\z + \frac12 )}.
\label{BBF}
\eea
The last term can be expressed in terms of a hypergeometric function as
\be
\frac{8 \sqrt{2} \pi}{v^{3/2} }\, \frac{\Gamma(\zeta + 1)}{\Gamma(\zeta+\frac32 )} \, 
 {}_3F_2(\frac32, \frac12, 1; 2, \zeta +\frac32 ;1).
\ee
For comparison with the $SU(N)$ result  (\ref{GammaF}), it is convenient to again express the hypergeometrics in terms of double integrals using (\ref{hyper}):
\be
\Gamma_F = 8 \, Q \, \left\{ - \, \frac{64}{3} \, \z \, x^{\frac32}y^{\frac12}\, \frac{(y-1)}{(x-1)} 
- 15  x^{\frac32}y^{\frac12}\,  \frac{(x-1)(y-1)}{(xy-1)}
+16 \frac{x y^{\frac32}}{(y-1)} \right\}.
\ee
The explicit $\z$ dependence can be eliminated from the first term by integrating by parts with respect to $y,$ allowing all three terms to be placed on a common denominator :
\bea
\Gamma_F &=&  8 \, Q \,  \frac{x^{\frac12} y^{\frac12}}{(x-1)(y-1)(xy-1)} \, \left\{ - \, 2 xy\,(1+xy) + 30 x^2 \, (1+xy) - 15 x \,(1+x^2y^2)   
\right.\nonumber \\ 
&& + \, \left. 16 x y^{\frac12} \, (1 + x^{\frac32}y^{\frac32}) - 16 xy^{\frac32} \, (1+x^{\frac12} y^{\frac12}) -11 x^2 y - 15 x^3 \right\}.
\eea
This is equivalent to the appropriately rescaled $SU(N)$ result (\ref{GammaF}), as the difference between the two can be expressed as
\bea
\frac{g^2}{\sqrt{ 2 \pi} v^{\frac32}} && \left[ \, 28 \, f( \frac12, \frac12, \frac52, 0, -1) -12 \, f( \frac12, \frac12, \frac52, 1, -1) \right. \nonumber \\
&& + \, \left.28 \, f( \frac12, \frac12, \frac52, 1, 0) \, \right],
\eea
which vanishes identically by (\ref{ident3}).

\sect{Conclusion}
In this paper, we have laid a basis for calculation of the two-loop effective action for the matrix model with gauge group $SU(N)$ in arbitrary bosonic backgrounds. This is important in extending tests of the BFSS conjecture to cases involving $N$ graviton scattering amplitudes.  We also achieve an important check of general results on the group theoretic structure of two-loop Feynman diagrams which have been previously used to compute sectors of the effective action for ${\cal N}=4$ supersymmetric Yang-Mills theory in four dimensions \cite{Kuzenko:2003wu,Kuzenko:2004ma,Kuzenko:2004sv}.

The two-loop effective action has been evaluated for a specific choice of background corresponding to scattering of a single D0-brane from a stack of $N$-1 D0-branes. The results are consistent with the $N$=2 results of  \cite{Becker:1998gp}, although establishing this requires some nontrivial identities. The need to introduce regularization did not arise, as all contributions calculated are manifestly finite.

For the background chosen, the  $N$-dependence of the two-loop effective action is in the form of an overall multiplicative factor $N(N-1).$ The agreement with the previously computed $N$=2 result \cite{Becker:1998gp} means that, for the particular background chosen,  
\be
\Gamma_{SU(N)} = \frac{N (N-1)}{2} \, \Gamma_{SU(2)}.
\ee
As a result, the structure of the two-loop contributions to the velocity dependent potential for the scattering of a single D0-brane from a stack of D0-branes is related to  the potential for scattering of two D0-branes by a simple scaling.  This provides  a specific test of the general argument presented in \cite{Becker:1997xw} suggesting that the two-loop contributions to the scattering potential should scale like $(N_1{}^2 N_2 + N_1 N_2{}^2)/2$ for $N_1$ D0-branes scattering from a stack of $N_2$ D0-branes.
\\

\noindent
{\bf Acknowledgements} \\ 
We are extremely grateful to Sergei Kuzenko for the impetus carry out this work, and for numerous discussions and suggestions. This work is supported in part by the Australian Research Council.

\begin{appendix}
\sect{Appendix}
This appendix outlines the derivation of the  expression (\ref{SUNkernel}) for the nonabelian kernel and its derivative (\ref{SUNkernelderiv}). We make use of the technique developed in   \cite{Kuzenko:2003eb}, building on earlier work for the coincidence limits of kernels in (\cite{McArthur:1997ww,Gargett:1998jq}). With gauge indices suppressed, the kernel satisfies the differential equation
\be
- \, \ri \frac{\rd }{\rd s} \,  K(\t,\t';s) = \Delta \, K(\t,\t';s)
\ee
with
\be 
\Delta = - (\partial_{\t} + \ri  g A_0 )^2 - g^2 \,\vec{Y}.\vec{Y} 
\ee
and with the boundary condition $\lim_{s \rightarrow 0} K(\t,\t';s) = \d(\t, \t'),$ for which the formal solution is 
\be 
K(\t,\t';s) = \re^{\ri s \Delta }\, \d(\t, \t') \, I(\t, \t').
\ee
Here, $I(\t, \t')$ is the parallel displacement propagator, which ensures the correct gauge transformation properties of the kernel at its endpoints.\footnote{For further details on the parallel displacement propagator and its properties, see   \cite{Kuzenko:2003eb}} Introducing a Fourier decomposition for the delta function, the kernel can be written
\be 
K(\t,\t';s) = \int_{- \infty}^{- \infty} \frac{\rd k}{2 \pi} \, \re^{\ri k (\t - \t')} \, \re^{\ri s \tilde{\Delta} }\, \d(\t, \t') \, I(\t, \t'),
\ee
where
\be
\tilde{\Delta} = - X_{\t}{}^2 - g^2 \, \vec{Y}.\vec{Y} 
\ee
with 
\be 
X_{\t} = \partial_{\t} + \ri g A_0 + \ri k.
\ee
The kernel satisfies the differential equation
\be 
-\, \ri \frac{\rd }{\rd s} \,  K(\t,\t';s) = -K_{\t \t}(\t,\t';s) -  g^2 \vec{Y}.\vec{Y} \, K(\t,\t';s),
\label{kerneldiff1}
\ee
where the ``moments'' $K_{\t \cdots \t}(\t,\t';s)$ (with $n$ subscripts $\t$) are  defined by 
\be
K_{\t \cdots \t}(\t,\t';s) = \int_{- \infty}^{- \infty} \frac{\rd k}{2 \pi} \, \re^{\ri k (\t - \t')} \, X_{\t}{}^n \,\re^{\ri s \tilde{\Delta} }\, \d(\t, \t') \, I(\t, \t').
\ee
The principle of the method developed in \cite{Kuzenko:2003eb,McArthur:1997ww} is to express the moment $K_{\t \t}(\t,\t';s)$ in terms of $K(\t,\t';s),$ so that (\ref{kerneldiff1}) becomes a linear differential equation for $K(\t,\t';s).$ This is relatively straightforward for simple backgrounds.

The identity
\be
0 = \int_{- \infty}^{- \infty} \frac{\rd k}{2 \pi} \, \frac{\rd}{\rd k}  \left( \re^{\ri k (\t - \t')} \, X_{\t} \, \re^{\ri s \tilde{\Delta} }\, \d(\t, \t') \, I(\t, \t') \right)
\ee
can be used to relate moments of different orders. Performing the differentiation 
yields the equation
\bea
0 &=&  \, (\t - \t') \, K_{\t }(\t,\t';s) + \, K(\t,\t';s) \nonumber \\
&& + \, 2 \sum_{0}^{\infty} \frac{ ( \ri s)^{n+1}}{(n+1)!}\int_{- \infty}^{- \infty} \frac{\rd k}{2 \pi} \, \re^{\ri k (\t - \t')} \, X_{\t} \, ad^{(n)}(\tilde{\D})(X_{\t}) \, \re^{\ri s \tilde{\Delta} }\, I(\t, \t').
\label{kerneldiff2}
\eea
For a covariantly constant background, $D_{\t}F_{\m \n}=0,$ 
\bea
 ad^{(2n+1)}(\tilde{\D})(X_{\t}) &=& (-1)^n \, 2^{2n+1} \, g^{2n+2} \,(D_{\t}\vec{Y}. D_{\t}\vec{Y})^n \, \vec{Y}. D_{\t}\vec{Y}, \quad n\geq 0 \nonumber \\
 ad^{(2n)}(\tilde{\D})(X_{\t}) &=& (-1)^n \, 2^{2n} \, g^{2n} \,(D_{\t}\vec{Y}. D_{\t}\vec{Y})^n \, X_{\t}, \quad n\geq 0.
\eea
Substitution into (\ref{kerneldiff2}) yields (with $\vec{F} = D_{\t} \vec{Y}$ and $F^2 = \vec{F}. \vec{F}$)
\bea
0 &=& (\t - \t') \, K_{\t }(\t,\t';s) + \cosh 2  g s F \,K(\t,\t';s) - \ri \, \frac{ \sinh  2 g s F}{gF} \, K_{\t \t}(\t,\t';s) \nonumber \\ && + \, ( \vec{Y}. \vec{F}) \, \frac{(\cosh 2 g s F -1)}{F^2} \,  K_{\t }(\t,\t';s).
\eea
This allows $K_{\t \t}(\t,\t';s)$ to be expressed in terms of $K(\t,\t';s),$ providing it is possible to express  $K_{\t}(\t,\t';s)$ in terms of $K(\t,\t';s).$

This is achieved by a similar method, namely using the identity
\be
0 = \int_{- \infty}^{- \infty} \frac{\rd k}{2 \pi} \, \frac{\rd}{\rd k}  \left( \re^{\ri k (\t - \t')} \,  \re^{\ri s \tilde{\Delta} }\, \d(\t, \t') \, I(\t, \t') \right),
\ee
which yields
\be
K_{\t }(\t,\t';s) = \frac{- \ri gF}{\sinh 2 g s F}\,  \left\{ (\t - \t') + (\cosh 2 g s F- 1) \, \left( \frac{\vec{Y}. \vec{F}}{F^2} \right) 
\right\} \, K(\t,\t';s).
\label{kerneldiff3}
\ee
Substituting back into (\ref{kerneldiff2}) then gives the required relation
\bea
\frac{\sinh 2 gs F}{g F} \,K_{\t \t}(\t,\t';s) &=& - \, \ri \, \cosh 2 g s F\, K(\t,\t';s) + \frac{gF}{\sinh 2 g s F} \, \left\{ (\t - \t') \right. \nonumber \\
&& + \, \left. (\cosh 2g sF - 1) \, \left( \frac{\vec{Y}. \vec{F}}{F^2} \right) \right\}^2 \, K(\t,\t';s).
\label{kerneldiff4}
\eea
In turn, substituting (\ref{kerneldiff4}) back into (\ref{kerneldiff1}) gives a linear first-order differential equation for $K(\t,\t';s)$ whose solution, subject to the boundary condition
\be
\lim_{s \rightarrow 0} \, K(\t, \t'; s) = \d(\t, \t'),
\ee
is 
\bea
 K(\t, \t'; s) &=& \left( \frac{\ri}{4 \pi s} \right)^{\frac12}  \re^{- \ri  g^2 s (Y^2 - F^2 (\frac{\vec{Y}. \vec{F}}{F^2})^2)} \, \left( \frac{2 g s F}{\sinh 2g s F} \right)^{\frac12}
\, \exp \left\{ - \, \ri   g F \frac{ (\t-\t')^2}{2} \,  \coth 2g s F \right. \nonumber \\&& + \, \left. \ri   g F \,  \left( \frac{\vec{Y}. \vec{F}}{F^2} \right) \, \left( \t - \t' -  \frac{\vec{Y}. \vec{F}}{F^2} \right) \,  \tanh gsF \right\}
 \, I(\t,\t').
\label{kerneldiff5}
\eea
Expression of the kernel in this form requires the identity
\be \frac{\cosh 2 g s F -1}{\sinh 2gsF} = \tanh g s F.
\ee

The derivative of the kernel follows immediately from (\ref{kerneldiff3}), as 
\be
D_{\t}  K(\t, \t'; s)= K_{\t }(\t,\t';s).
\label{kerneldiff6}
\ee
Note that it is not trivial to compute $D_{\t}  K(\t, \t'; s)$ simply by taking the covariant derivative of the expression (\ref{kerneldiff5}) for $ K(\t, \t'; s),$ as the derivative of the parallel displacement propagator $I(\t, \t')$ must be computed. Making use of (\ref{kerneldiff6}) neatly bypasses the need to compute this derivative. 

Using the above approach, it is very easy to see the difference between carrying out quantum calculations in ten dimensions and then dimensionally reducing to 1+0 dimensions, as opposed to carrying out quantum calculations directly in the dimensionally reduced theory. After Fourier decomposition of the delta function, the ten-dimensional heat kernel for the operator 
\be
\Delta_{10} = - \, (\partial_{\t} + i g A_0)^2 + (\partial_i + i g Y_i)(\partial_i + i g Y_i), \quad i = 1, 2, \cdots, 9
\ee
is
\bea
K_{10}(x,x\,' ; s) &=& \int \frac{\rd^{10}k}{(2 \pi )^{10}} \, \re^{i k_0(\t - \t') + i k_i (x-x')_i } \nonumber \\
&& \times \, \re^{i s (- \,X_{\t}^2 +  (\partial_i + i g Y_i +i k_i) (\partial_i + i g Y_i +i k_i))}.
\eea
Dimensionally reducing by setting $x_i = x'_i$ and making all fields independent of $x_i$ (so that the $\partial_i$ do not contribute), the result is 
 \bea
K_{10}(x,x\,' ; s) &=& \int \frac{\rd k_0}{2 \pi } \, \re^{  i k_0(\t - \t') } \, \int \frac{\rd^9 k}{(2 \pi)^9} \, \re^{i k_i (x-x')_i } \nonumber \\
&& \times \, \re^{- i s k_i k_i} \, \re^{i s ( - X_{\t}^2  - g^2 Y_i Y_i - 2 g k_i Y_i) }.
\label{K10}
\eea
On the other hand, the kernel for the dimensionally reduced operator
\be
\Delta =  - \, (\partial_{\t} + i g A_0)^2 - g^2 Y_i Y_i
\ee
is
 \be
K(x,x\,' ; s) = \int \frac{\rd k_0}{2 \pi } \, \re^{ i k_0(\t - \t') }
 \, \re^{i s (- X_{\t}^2  - g^2 Y_i Y_i ) }.
 \label{K0}
\ee
The inability to factor the term $k_i Y_i$ out of the last exponential in (\ref{K10}) without generating additional structure via the  Baker-Campbell-Hausdorff formula means that the relationship between the kernels (\ref{K10}) and (\ref{K0}) is highly nontrivial.

\sect{Appendix}
In this appendix, we prove, under certain assumptions about the nature of the background, that two terms in the contribution (\ref{GammaIII0}) to the effective action with different Lorentz structures are proportional. These are
\be
 - \, \frac{\ri g^2 }{2} \, \int \rd \t \, G^{ \m a \, \n b } (\t,\t') 
   \,   \tr_{\rm A} \left( T_a \, D_{\m}  G_{\s \n}(\t,\t') \, T_b \, D\,'_{\rho} G^{\r \s}(\t',\t) \right)  = - \, C_1
\ee
and 
\be - \,\frac{3 \ri g^2 }{2} \, \int \rd \t \, G^{ \m a \, \n b } (\t,\t') 
 \, \tr_{\rm A} \left( T_a \, D_{\m}  G^{\s \r}(\t,\t') \, T_b \, D\,'_{\r}G_{\n \s}(\t',\t) \right) = - \, 3 \, C_2.
 \ee
 Specifically, we will show $C_1 = C_2$ when the $SU(N)$ background is of the form 
 \be
 g A_{\m}(\t) = (0, \vec{v} \t + \vec{b}) \, H,
 \label{choice}
 \ee
  with $H$ an arbitrary element of the Cartan subalgebra. This covers scattering of arbitrary configurations of $N$ D0-branes, and not just the special case considered in detail in this paper.
 
The first step is to use the factorized form (\ref{factoredvector}) for the vector Green's function and the covariant constancy of the field strength $F_{\m \n}$ to write
\bea
C_1 - C_2 &&= \frac{\ri  g^2 }{2} \, \int \rd \t \, G^{ \m a \, \n b } (\t,\t') \, \ri^2 \, \int_0^{\infty} \rd s \int_0^{\infty} \rd t \, \nonumber \\ && \times \, \left\{
   \,   \tr_{\rm A} \left( T_a \, (\re^{-2 s g F})_{\s \n} \, D_{\m}  K(\t,\t' ;s) \, T_b \,  (\re^{-2 t g F})^{\r \s} \, D\,'_{\rho}  K(\t',\t ;t)  \right) \right.  \nonumber \\ && - \, \left.
\tr_{\rm A} \left( T_a \, (\re^{-2 s g F})^{\s \r} \, D_{\m}  K(\t,\t' ;s) \, T_b \,  (\re^{-2 t g F})_{\n \s} \, D\,'_{\rho}  K(\t',\t ;t)  \right) \right\}
\eea
If the background fields belong to the Cartan subalgebra, it is possible to change the order of the first exponential and the first kernel in each trace, so that it suffices to prove that
\be  
C = (\re^{-2 s g F})_{\s \n} \,  T_b \,  (\re^{-2 t g F})^{\r \s} -  (\re^{-2 s g F})^{\s \r} \,  T_b \,  (\re^{-2 t g F})_{\n \s}
\ee
vanishes. Writing $F_{\m \n} = f_{\m \n} \, H,$ it follows that
\be 
(\re^{-2 s g F})_{\s \n} \,  T_b = T_b \, (\re^{-2 s g f (H + e_b)})_{\s \n},
\ee
where $e_b$ is the $U(1)$ weight of the generator $T_b$ with respect to $H,$
\be 
[H, T_b ] = e_b \, T_b.
\ee
As a result,
\be  
C = T_b \,  (\re^{-2  g f(t H + s H + e_b)})^{\r}{}_{ \n} -  T_b \,  (\re^{-2  g f(t H + s H + e_b)})_{\n}{}^{\r}.
\ee
This vanishes, since with the choice (\ref{choice}) of background, $f^{\r}{}_{ \n}$ is a symmetric matrix,
\be
f^{\r}{}_{ \n} = \left( \begin{array}{cc} 0 & \quad - \, \vec{v}^{\, T} \\ - \, \vec{v} & \quad {\bf 0} \end{array} \right).
\ee

\end{appendix}

\end{document}